\documentclass[acmtog]{acmart}
\usepackage{algorithm}
\usepackage{algorithmic}
\usepackage{subfigure}

\AtBeginDocument{%
  }

\copyrightyear{2025} 
\acmYear{2025} 
\setcopyright{acmlicensed}\acmConference[SIGGRAPH Conference Papers '25]{Special Interest Group on Computer Graphics and Interactive Techniques Conference Conference Papers }{August 10--14, 2025}{Vancouver, BC, Canada}
\acmBooktitle{Special Interest Group on Computer Graphics and Interactive Techniques Conference Conference Papers (SIGGRAPH Conference Papers '25), August 10--14, 2025, Vancouver, BC, Canada}
\acmDOI{10.1145/3721238.3730720}
\acmISBN{979-8-4007-1540-2/2025/08}

\acmSubmissionID{955}

\begin{document}

\title{MGPBD: A Multigrid Accelerated Global XPBD Solver}

\author{Chunlei Li}
\authornote{Both authors contributed equally to this research.}
\email{li_cl@foxmail.com}
\orcid{0000-0002-9822-9465}
\author{Peng Yu}
\email{yupeng@buaa.edu.cn}
\authornotemark[1]
\orcid{0000-0002-8652-2744}
\affiliation{%
  \institution{State Key Laboratory of Virtual Reality Technology and Systems, Beihang University}
  \city{Beijing}
  \country{China}
}

\author{Tiantian Liu}
\email{ltt1598@gmail.com}
\orcid{0000-0002-5614-1588}
\affiliation{%
  \institution{Taichi Graphics}
  \city{Beijing}
  \country{China}
}

\author{Siyuan Yu}
\email{ysysimon@live.com}
\orcid{0009-0003-1310-0094}
\affiliation{%
  \institution{Zenustech}
  \city{Shenzhen}
  \country{China}
}

\author{Yuting Xiao}
\email{1035494078@qq.com}
\orcid{0009-0001-6480-2753}
\affiliation{%
  \institution{State Key Laboratory of Virtual Reality Technology and Systems, Beihang University}
  \city{Beijing}
  \country{China}
}

\author{Shuai Li}
\email{lishuai@buaa.edu.cn}
\orcid{0000-0003-4182-1588}
\authornote{Corresponding authors.}
\affiliation{%
  \institution{State Key Laboratory of Virtual Reality Technology and Systems, Beihang University}
  \city{Beijing}
  \country{China}
}

\author{Aimin Hao}
\email{ham@buaa.edu.cn}
\orcid{0000-0002-5774-6706}
\affiliation{%
  \institution{State Key Laboratory of Virtual Reality Technology and Systems, Beihang University}
  \city{Beijing}
  \country{China}
}

\author{Yang Gao}
\authornote{Corresponding authors.}
\email{gaoyangvr@buaa.edu.cn}
\orcid{0000-0002-9149-3554}
\affiliation{%
  \institution{State Key Laboratory of Virtual Reality Technology and Systems, Beihang University}
  \city{Beijing}
  \country{China}
}

\author{Qinping Zhao}
\email{zhaoqp@buaa.edu.cn}
\orcid{0000-0001-5600-5300}
\affiliation{%
  \institution{State Key Laboratory of Virtual Reality Technology and Systems, Beihang University}
  \city{Beijing}
  \country{China}
}

\begin{abstract}
We introduce a novel Unsmoothed Aggregation (UA) Algebraic Multigrid (AMG) method combined with Preconditioned Conjugate Gradient (PCG) to overcome the limitations of Extended Position-Based Dynamics (XPBD) in high-resolution and high-stiffness simulations. While XPBD excels in simulating deformable objects due to its speed and simplicity, its nonlinear Gauss-Seidel (GS) solver often struggles with low-frequency errors, leading to instability and stalling issues, especially in high-resolution, high-stiffness simulations. Our multigrid approach addresses these issues efficiently by leveraging AMG. To reduce the computational overhead of traditional AMG, where prolongator construction can consume up to two-thirds of the runtime, we propose a lazy setup strategy that reuses prolongators across iterations based on matrix structure and physical significance. Furthermore, we introduce a simplified method for constructing near-kernel components by applying a few sweeps of iterative methods to the homogeneous equation, achieving convergence rates comparable to adaptive smoothed aggregation (adaptive-SA) at a lower computational cost. Experimental results demonstrate that our method significantly improves convergence rates and numerical stability, enabling efficient and stable high-resolution simulations of deformable objects.
\end{abstract}



\begin{teaserfigure}
  \includegraphics[width=\textwidth]{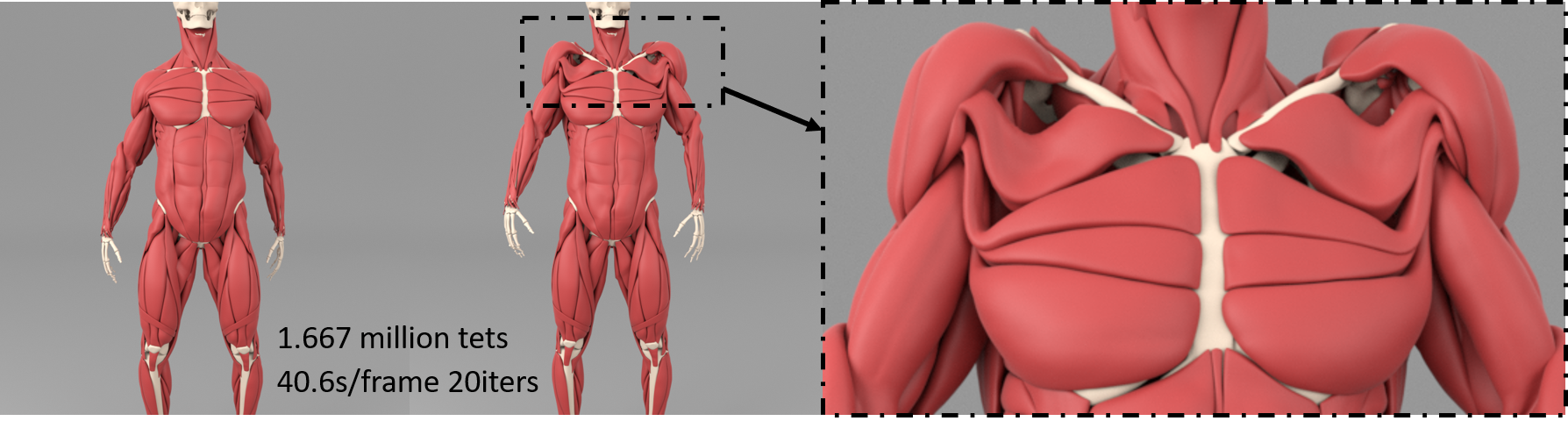}
  \caption{Human muscle case (1.667M tetrahedrons) simulated via MGPBD. 20 iterations per frame. Performance: 40.6 s/frame. }
  \Description{}
  \label{fig:teaser}
\end{teaserfigure}

\maketitle

\section{Introduction}

Efficient, precise, and stable deformable object simulations are essential for numerous applications such as games, films, etc.  The Position-Based Dynamics (PBD) method  and the Extended PBD (XPBD) method ~\cite{Macklin2016-XPBD}  have gained widespread use for simulating rigid bodies, soft bodies, and fluids within a limited computational budget due to their simplicity and robustness ~\cite{Muller2007-PBD}.

However, both PBD and XPBD experience significant convergence issues in practical scenarios, especially when dealing with high-resolution meshes, large time steps, or extremely stiff materials. Under such conditions, simulations often struggle to stabilize, even after thousands of iterations, frequently leading to non-physical artifacts or complete solver failures.Wang et al.~\cite{Wang2015-Chebyshev} proposed the Chebyshev method to enhance the convergence of PBD, but its effectiveness is limited. 
This is because the iterative solver  PBD/XPBD adopted efficiently reduces high-frequency errors but has difficulty with low-frequency errors. Additionally, the PBD/XPBD iterative solver omits the off-diagonal terms of the system matrix, which relate to adjacent constraints with shared vertices, further constraining its  globality.

In contrast, the multigrid (MG) method is effective in addressing both high-frequency and low-frequency errors. Inspired by MG,  \citet{Muller2008-HPBD} proposed the hierarchical PBD (HPBD) method and  \citet{multilayerxpbd} introduced multilayer PBD, but being multi-resolution rather than multigrid methods, they may struggle to ensure physical consistency between levels.

In addition, direct implementation of conventional multigrid methodologies may fail to yield meaningful performance gains within the XPBD framework.  More critically, the unmodified application of multigrid techniques to XPBD's global system induces numerical instability, leading to solution divergence. Synergistic optimization of the smoothing and prolongation operators is key to optimal efficiency. Moreover, reconstructing the multigrid hierarchy, including generating coarse grids and recalculating interpolation operators, is required in each iteration. This setup step incurs the primary computational cost, and it consumes the majority of the solver's total time.

Previous studies \cite{Xian2019-GalerkinMG,Ruan2024MiNNIE}  applied AMG to accelerate deformable body simulations. Unlike these primal-space-based works, our research pioneers the use of AMG in the dual space. As \citet{Macklin2020-primal-dual} pointed out, dual-space methods are superior in handling high stiffness-to-weight ratios, crucial for muscle simulations, whereas primal methods fail.

This study introduces MGPBD (Multigrid Position Based Dynamics), a novel Unsmoothed Aggregation (UA) Algebraic Multigrid (AMG) Preconditioned Conjugate Gradient (PCG) solver for the global XPBD system. We improved the near-kernel components, ensuring convergence and boosting multigrid performance. By leveraging the constant matrix structure across iterations when topology remains unchanged, a lazy update strategy defers the costly setup phase, significantly reducing computational overhead. Additionally, the unsmoothed aggregation approach enhances coarse-grid matrix sparsity, minimizing the cost of sparse matrix-vector multiplication operations and mitigating computational bottlenecks.
Experiments demonstrate MGPBD's superiority over XPBD, achieving stable simulations in high-resolution and high-stiffness scenarios (Fig.~\ref{fig:teaser}, Fig.~\ref{fig:monster}). The source code is available at \url{ https://github.com/chunleili/mgpbd} 
\section{Related Work}

The simulation of deformable objects began with Terzopoulos \cite{Terzopoulos1987}, who first applied the finite element method (FEM) to the computer graphics community. Baraff et al. \cite{Baraff-Witkin1998-LargeStep} proposed an implicit time integration method for cloth, widely used in large-time-step soft body simulations \cite{kimdeformables, KimBWCloth}. Some research  \cite{Bouaziz2014-PD, li2022energetically, Lan2023-2nd-order-interior-point, Chen2024-VBD}  reformulated dynamic problems as optimizations via variational Euler methods. In all of these
works, the use of an implicit integrator that requires a global linear
solver is commonplace.

Due to its simplicity and efficiency, PBD is widely used in the simulation of soft bodies, cloth, rigid bodies, and fluid~\cite{Macklin2016-XPBD, Macklin2021-DiffSim, Bender2017-Survey-PBD,gao2019efficient,Macklin2021-Neohooken-PBD}. However, PBD and XPBD often face convergence problems. Many studies tried to improve PBD convergence. \citet{Macklin2019-small-steps} found that substepping with a small timestep and a single constraint solver iteration is more stable. Adding geometric stiffness boosts stability but raises costs.  \citet{Chen2023-PrimalPBD, ChenPBD2024Siggraph} pointed out that excluding primary equations in large gradient changes hinders residual reduction, while including them may cause instabilities.

MG methods are famous for their rapid convergence, particularly when dealing with large meshes and detail-rich deformations, making them popular in the computer graphics community. Building prolongators is a major challenge in  multigrid methods. The geometric multigrid relies on carefully predefined multi-resolution meshes to address this challenge. For instance, \citet{Georgii-Westermann2006-multigrid} employed a nested cage of tetrahedrons; \citet{zhu2010-Multigrid-Solid} and \citet{Dick2011-MG-CUDA} constructed prolongators based on a background hexahedral mesh; \citet{wang2020-hierarchical-mpm} applied the multigrid technique to the Material Point Method (MPM), where the background hexahedral mesh is already inherent; and \citet{liu2021-surface-multigrid} built prolongators on a surface mesh through successive self-parameterization.  Unlike GMG, AMG is agnostic of geometric tessellation and automatically constructs multi-level hierarchies solely from the system matrix. \citet{Tamstorf2015-SA-Disney} designed a smoothed aggregation for cloth simulation. \citet{Li2023-subspace-cloth} combined MG with reduced-order models for contact-aware cloth. \citet{Xian2019-GalerkinMG}proposed an interpolation technique using linear blend skinning, which adeptly manages high-resolution meshes while bypassing the necessity for intensive linear resolution computations. \citet{Ruan2024MiNNIE} proposed the MiNNIE approach based on the mixed FEM framework to accelerate nonlinear near-incompressible elastic objects and used the same prolongator as \citeauthor{Xian2019-GalerkinMG}. 

In conclusion, there are currently no AMG-based XPBD solvers within the computer graphics community. Traditional AMG methods require a complete reconstruction of grid hierarchies, including prolongation operators and sparse system matrices, whenever the values of the matrix elements change, a process that accounts for the majority of the computational time per iteration. Designing a multigrid algorithm that achieves full computational efficiency for new applications remains a challenging task. To this end, our aim was to leverage the physical characteristics of the XPBD linear system to develop an AMG-based approach that enables stable and efficient simulations of high-resolution soft bodies.

\section{Revisiting XPBD}
The dynamics of a system could be formulated as a quadratic minimization problem under the XPBD framework:
\begin{equation}
    \mathbf{x}^{t+1} = \operatorname*{argmin}_{\mathbf{x}} (U( \mathbf{x} ) + \frac{1}{2 \Delta t^2}|| \mathbf{x} -\tilde{ \mathbf{x} }||^2_{\mathbf{M}}) \label{eq:energy-minimization}, 
\end{equation}
where $\mathbf{x} \in \mathbb{R}^{3\times N}$ is the position of $N$ particles that represent the deformable objects, $\Delta t$ is the timestep size, $U(\mathbf{x}) = \frac{1}{2} \mathbf{C(x)}^{\top} \boldsymbol{\alpha}^{-1} \mathbf{C(x)}$ is the quadratic potential energy consisting of a compliant constraint function $\mathbf{C(x)} \in \mathbb{R}^{m}$, $m$ is the number of constraints, $\mathbf{M}$ is the diagonal mass matrix, $\boldsymbol{\alpha}$ is the compliance matrix, $\tilde{\mathbf{x}} = \mathbf{x}^t + \Delta t\mathbf{v}^{t+1} $ is the predicted position, superscript $t$ and $t+1$ denote the time steps, which are omitted in the following for simplicity.
The first-order optimality condition reduces the problem to a nonlinear system. Through local linearization (e.g., Newton-Raphson), this system is reformulated as a linear problem:

\begin{equation}
    \begin{bmatrix}
        \mathbf{M} & -\nabla C^{\top} \\
        \nabla C & \boldsymbol{\tilde{\alpha}}
    \end{bmatrix}
     \begin{bmatrix}
        \Delta \mathbf{x} \\
        \Delta \boldsymbol{\lambda}
    \end{bmatrix} = 
    \begin{bmatrix}
        \mathbf{0} \\
        -\boldsymbol{C(\mathbf{x})} -\tilde{\boldsymbol{\alpha}} \boldsymbol{\lambda}
    \end{bmatrix},\label{eq:Newton-step}
\end{equation}
where $\boldsymbol{\lambda}=-\tilde{\boldsymbol{\alpha}}^{-1} \boldsymbol{C}(\mathbf{x})$ and $\tilde{\boldsymbol{\alpha}} = \boldsymbol{\alpha} / \Delta t^2$.  $\boldsymbol{h}(\mathbf{x}, \boldsymbol{\lambda}) = \boldsymbol{C}(\mathbf{x}) + \tilde{\boldsymbol{\alpha}}\boldsymbol{\lambda}$ represents the dual residual of the Lagrange multiplier unknowns.
Applying the Schur complement with respect to $\mathbf{M}$ leads to a reduced linear system 
\begin{equation}
\underbrace{(\nabla \boldsymbol{C(\mathbf{x})} \mathbf{M}^{-1} \nabla \boldsymbol{C(\mathbf{x})}^{\top} + \boldsymbol{\tilde{\alpha}})}_{\mathbf{A}} \Delta \boldsymbol{\lambda}=\underbrace{- \boldsymbol{C(\mathbf{x})} - \tilde{\boldsymbol{\alpha}} \boldsymbol{\lambda}}_{\mathbf{b}}\label{eq:Ax=b}.
\end{equation}
The linear equation Eq.~\ref{eq:Ax=b} with a system matrix $\mathbf{A}$ can be solved by an iterative solver such as the GS or Jacobi solver. The XPBD method employs a more simplified version of the GS/Jacobi solver, retaining only the diagonal elements of the linear matrix to sequentially compute the change of the Lagrange multiplier for each constraint:
\begin{equation}
    {\Delta \lambda_j} = A_{j,j}^{-1} b_j = (\nabla {C_j} \mathbf{M}^{-1} \nabla {C_j}^{\top} + {\tilde{\alpha_j}})^{-1}(- {C}_j - \tilde{{\alpha}_j} \lambda_j).
\end{equation}

The omission of off-diagonal terms that capture adjacent information confines XPBD to being a local solver, thereby hindering its ability to propagate errors effectively. In contrast, \citet{Goldenthal2007-SAP} solved the same linear system using a direct solver with hard constraints, setting the compliance as zero. However, the direct solver's cubic time complexity $O(n^3)$ and excessive memory consumption for large-scale sparse matrices render this method computationally prohibitive for high-resolution objects. Finally, the position correction $\Delta \mathbf{x}$ is calculated by a substitution of $\Delta \boldsymbol{\lambda}$ into Eq.~\ref{eq:Newton-step}, which gives: 
\begin{equation}
    \Delta \mathbf{x} = \mathbf{M}^{-1} \nabla \boldsymbol{C(\mathbf{x})}^{\top} \Delta \boldsymbol{\lambda}.
\end{equation}
\section{MGPBD}
 
\begin{figure}
    \centering
    \includegraphics[width=\the\linewidth]{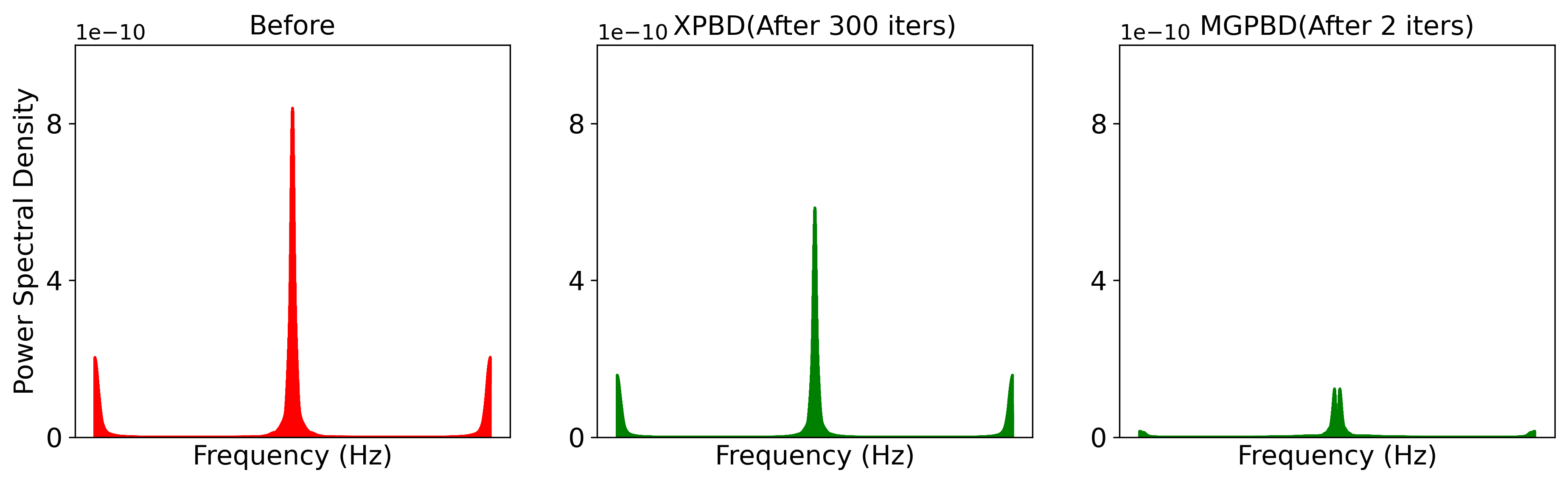}
    \caption{Power density graph after 2 iterations of MGPBD and 300 iterations of XPBD. From left to right, the red subgraph shows dual residuals before the solving, while the other green ones are after solving. Two iterations of MGPBD eliminate the two spikes at LF and HF, while 300 iterations of XPBD solving still keep the spikes. This explains the superiority of MG in solving all-frequency errors over the iterative method. The data is generated by a cloth hanging case consisting of $1025\times1025$ particles at the 11th frame.}
    \Description{Power density graph after 2 iterations of MGPBD and 300 iterations of XPBD.}
    \label{fig:power_density}
\end{figure}

We believe that the convergence issue is primarily due to the simplified iterative solver of XPBD. Firstly, the locality of the iterative solver hinders error propagation, preventing the resolution of low-frequency errors. Secondly, the omission of off-diagonal terms exacerbates this locality problem. In contrast, a multigrid solver can efficiently eliminate both low-frequency and high-frequency errors. Fig.~\ref{fig:power_density} illustrates that the residual power spectral density is effectively reduced after two iterations of MGPBD, while 300 iterations of XPBD fail to achieve the same outcome.

Therefore, in this study, we propose an MGPBD model that uses AMG to solve the global system of Eq.~\ref{eq:Ax=b}. The MGPBD simulation loop is shown in Algo.~\ref{algo:substep-mgpbd}. Compared to XPBD, the key distinction lies in the assembly of the global system matrix and the solution of the linear system using the MGPCG solver. Assemble the system matrix $\mathbf{A}$ (line 5) following the procedure outlined in Sec. \ref{sec:system-matrix}. The setup phase (Sec. \ref{sec-setup}) is lazily updated every few frames (e.g., 20) (line 7), leveraging the invariance of the matrix structure when the mesh topology remains unchanged. The linear system is then solved using the MGPCG solver (line 8). Subsequently, the position correction is computed and the Lagrange multipliers and positions are then updated (lines 9–11). 

The $\omega$ in line 11 is a relaxation factor that can be either user-specified (e.g., 0.1 for softbody and 0.25 for cloth) or automatically determined through a backtracking mechanism that halves $\omega$ when the residual increases.

\begin{algorithm}[t]
    \begin{algorithmic}[1]
    \caption{MGPBD Simulation Loop}
    \label{algo:substep-mgpbd}
    \STATE $\tilde{\mathbf{x}}$,$\mathbf{x}$,$\mathbf{x_{old}, \mathbf{v}}$
 $\leftarrow$ semiEuler($\mathbf{v}$,$\Delta t$,$\mathbf{f_{ext}}$)
    \STATE $\boldsymbol{\lambda}\leftarrow(0,...,0)^T$
    
    \FOR {$ite=0,1,...,maxiter$}
        \STATE calculate  $\mathbf{C}$ and $\nabla C$
        \STATE assemble $\mathbf{A} \leftarrow \nabla C \mathbf{M}^{-1} \nabla C^{\top} + \boldsymbol{\tilde\alpha}$
        \STATE calculate $\mathbf{b} \leftarrow - \mathbf{C} - \boldsymbol{\tilde\alpha} \boldsymbol{\lambda}$
        \STATE setup AMG for every few frames (e.g., 20).
        \STATE solve $\mathbf{A} \Delta \boldsymbol{\lambda} = \mathbf{b}$ using MGPCG solver
        \STATE $\Delta \mathbf{x} \leftarrow \mathbf{M}^{-1} \nabla C^{\top} \Delta\boldsymbol{\lambda}$
        \STATE $\boldsymbol{\lambda} \leftarrow \boldsymbol{\lambda} + \Delta\boldsymbol{\lambda}$
        \STATE $\mathbf{x} \leftarrow \mathbf{x} + \omega \Delta \mathbf{x}$
        \IF{timeBudgetExhausted or $\|\mathbf{b}\| < \epsilon$}
            \STATE break
        \ENDIF
    \ENDFOR
    \STATE collision response
    \STATE $\mathbf{v}\leftarrow (\mathbf{x}- \mathbf{x_{old}})/\Delta t$
\end{algorithmic}
\end{algorithm}

\begin{figure*}
    \centering
    \includegraphics[width=\the\linewidth]{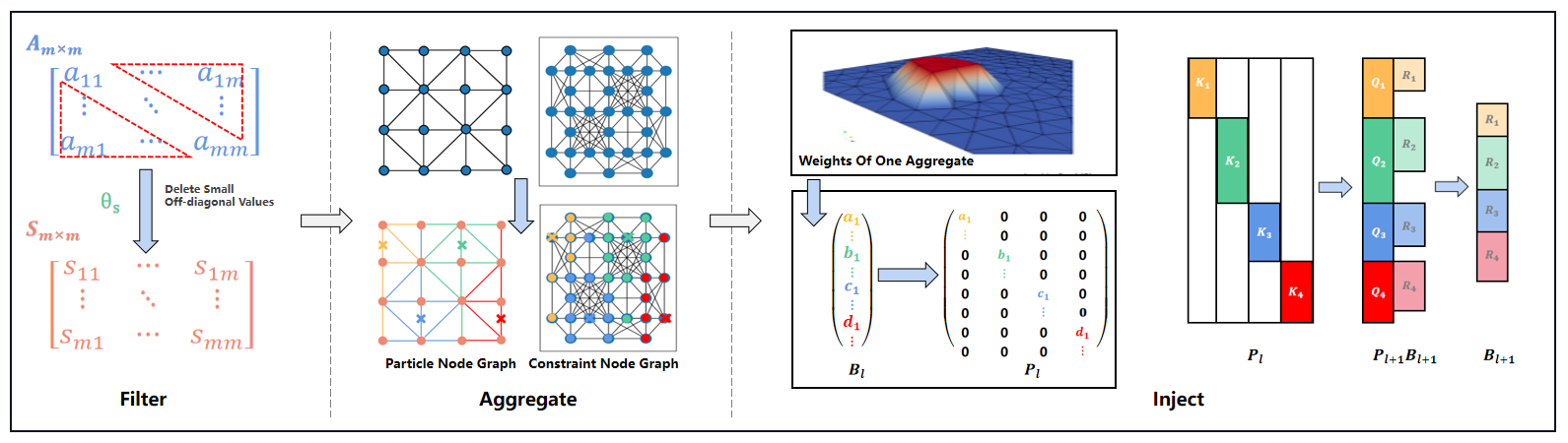}
    \caption{ Pipeline of the setup phase of UAAMG. Input: $A$ at the finest level; Output: $P$s at all levels. It includes 3 steps: 1) {Filter}: filter out the weak connection with a threshold $\theta_s$ and get the strength of connection matrix $\mathbf{S}$. Here $\mathbf{A}$ is symmetric, and off-diags $|a_{ij}|$ are used as strength. 2) {Aggregate}: aggregate the nodes into separate groups using $\mathbf{S}$. Here is an example of cloth with $4\times4$ particles. Its dual graph (constraint node graph) is aggregated into four groups marked in yellow, green, blue, and red, respectively. 3) {Inject}: inject a near-kernel component $\mathbf{B}$ into the matrix $\mathbf{P}$. Each column of $\mathbf{P}$ corresponds to one aggregate, and its nonzero value corresponds to the weight within that aggregate. For every level, use QR decomposition to form the next level, where $\mathbf{R}$ of QR decomposition serves as $\mathbf{B}$ at the next level. With $\mathbf{P_l}$ at this level, the next level $A$ will be formed with Eq.~\ref{eq:Galerkin} and used as the input to construct the next level $\mathbf{P_{l+1}}$. This procedure is repeated for all levels until the matrix size is under $400$. We finally get a set of $\mathbf{P}$s as the output of the AMG setup phase.}
    \Description{Pipeline of setup phase of UAAMG.}
    \label{fig:setup-pipline}
\end{figure*}

\subsection{Setup Phase}\label{sec-setup}
The AMG solver consists of two phases: the setup phase and the solving phase. The setup phase focuses on the construction of the prolongators, while the solving phase involves solving the linear system using these prolongators.

The main steps of the setup phase, illustrated in Fig.~\ref{fig:setup-pipline}, include three key steps: 1) \textbf{Filter}: Constructing the strength-of-connection matrix, 2) \textbf{Aggregate}: Aggregating the nodes into several interleaved groups, and 3) \textbf{Inject}: Computing the prolongator through injection. The first step involves creating a Strength-Of-Connection (SOC) matrix $\mathbf{S}$ based on the relative off-diagonal values. This step aims to eliminate weak connections by $|A_{ij}| < \theta_s \cdot \sqrt{|A_{ii}||A_{jj}|}$, where $\theta_s$ is a user-defined strength threshold that impacts the convergence rate and performance. Our findings indicate that a value of 0.1 is optimal for most softbody cases. Then we employ the standard aggregation algorithm \cite{Tamstorf2015-SA-Disney}, beginning by selecting a node and aggregating it with strongly connected neighbors. The next unaggregated neighbor node forms the second aggregate. This process repeats until all nodes are in an aggregate or left alone. To handle the unaggregated nodes, we assign them to the neighbor aggregate with the strongest connection.
After aggregation, the interleved aggregates are stored in an array representing the aggregate index to which the nodes belong. The sparsity pattern of the prolongator $\mathbf{P}$ is formed such that each column corresponds to one aggregate, with the nonzero values representing the weights within that aggregate. Finally, in the "inject" step, the near-kernel vector $\mathbf{B}$ is injected into the prolongator $\mathbf{P}$. The injected values correspond to the interpolation weights assigned to each node within a given aggregate. Subsequently, the near-kernel components for the next layer, $\mathbf{B}_{l+1}$, are computed by QR factorization of the near-kernel components of the current layer, $\mathbf{B}_l$,
where $j$ denotes the $j-$th constraint.

\begin{figure}
    \centering
    \includegraphics[width=1.0\linewidth]{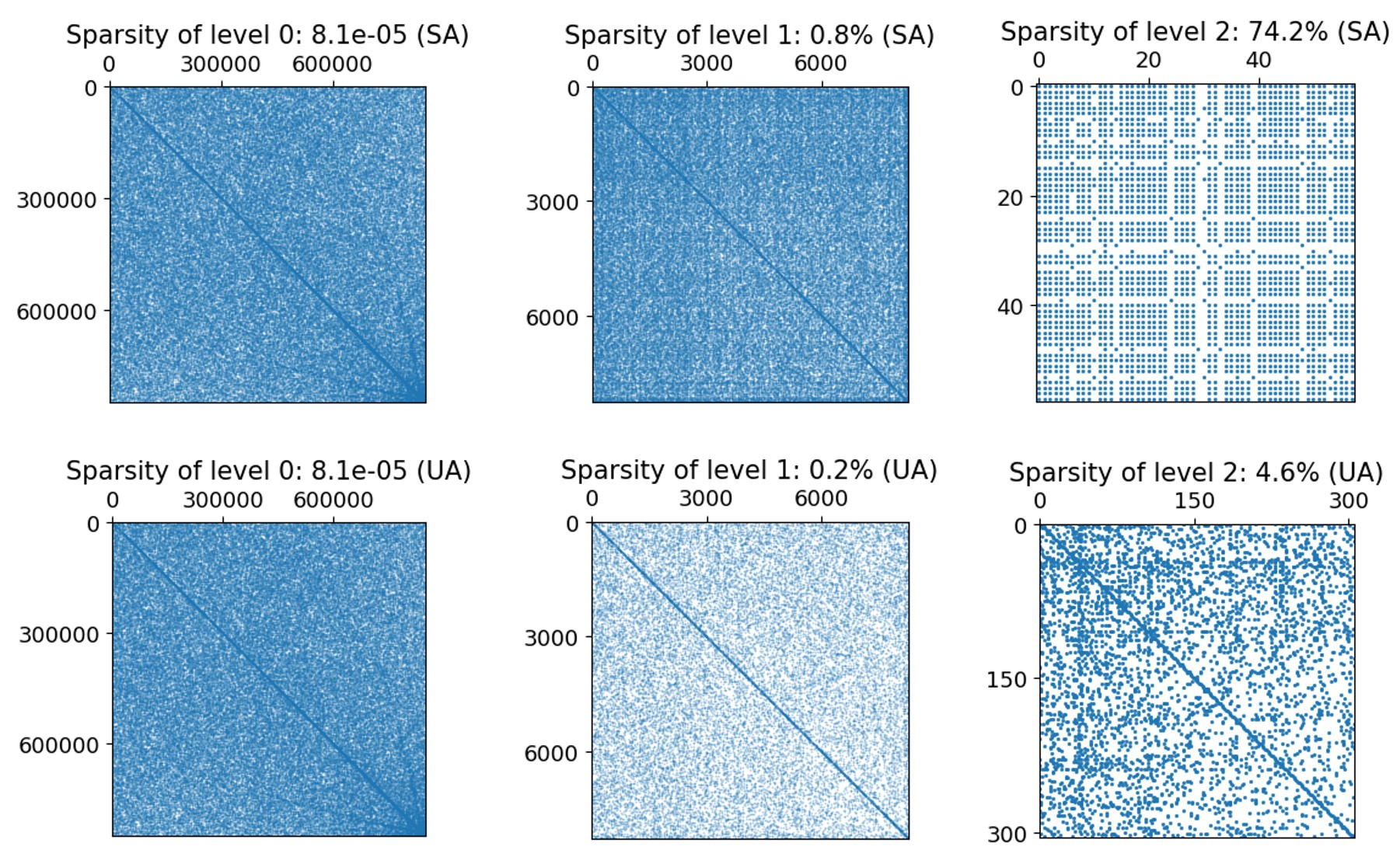}
    \caption{Sparsity patterns of SA and UA. Number of nonzeros ($nnz$): 58M/587K/2K (SA) vs. 58M/136K/4K (UA). Sparsity: 8e-5/0.8\%/74.2\% (SA) vs. 8e-5/0.2\%/4.6\% (UA) VS. Data is collected from the first frame of bunny squash case with 850K tetrahedrons. The key difference of $nnz$ in level 1 makes UA faster than SA.}
    \Description{Sparsity patterns of matrices of SA and UA.}
    \label{fig:sparsity}
\end{figure}

We adopt Unsmoothed Aggregation (UA), which does not perform the smoothing prolongator steps as Smoothed Aggregation (SA) \cite{Vanek1996-SA}. The reason is that the smoothing prolongator step can lead to denser coarse-grid matrices, which may significantly increase computational overhead \cite{Xian2019-GalerkinMG}.  This approach yields a sparser coarse-grid matrix (Fig.~\ref{fig:sparsity}), and thus makes the UA faster than SA while achieving a nearly identical convergence rate (Fig. \ref{fig:compare-linearSolver}).

\subsection{Lazy Setup Strategy}\label{sec:system-matrix}
Through our observations, we find that at every iteration step, the setup phase consumes most (up to two-thirds) of the overall computational duration. To enhance the performance in solving linear systems, we propose a lazy setup strategy. Next, we will explain the rationale behind this approach. For example, a softboy represented as a tetrahedral mesh with cardinality-4 constraints, the diagonal of $\mathbf{A}$ is $\mathbf{A}_{ii}=\sum_{k=0}^3 m_{i_k}^{-1} \mathbf{g}_{i_k}^2+\tilde{\alpha}$, and the off-diagonal terms are $\mathbf{A}_{ij}=\sum_{k=0}^3 m_{sv}^{-1} \mathbf{g}_{i_k}\cdot\mathbf{g}_{j_k}$, where $i, j$ denote the indices of two adjacent elements, $k$ represents the $k$th vertex of an element, $sv$ refers to the shared vertex, and $\mathbf{g}\in \mathbb{R}^3$ denotes the gradient with respect to a vertex. $\mathbf{A}_{ij}$ undergoes minimal changes as long as there are no significant variations in its gradients or the angles between them. Specially, in the system with distance constraints, $\mathbf{A}_{ii}$ remains constant. Therefore, the setup phase dependent on the matrix pattern and the relative values of its nonzero elements can be updated lazily.  Additionally, the matrix can be precomputed and stored in Compressed Sparse Row (CSR) format with three fixed length arrays. For each row, we store the off-diagonal terms first and put the diagonal terms at last. This fixed structure further improves performance.

\begin{figure}[t]
    \centering
    \includegraphics[]{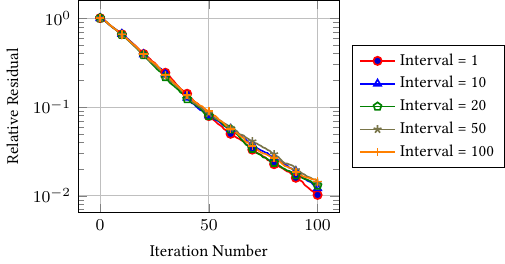}
\caption{
    Residual curves for different setup intervals. There is no significant difference in convergence when updating the prolongators aross $1/10/20/50/100$ frames. Data are collected from the $99$-th frame of the $270$K-resolution bunny squash scenario (Fig. \ref{fig:bunny}).
}
\Description{A line plot to show the lazy setup has no impact on convergence.}
    \label{lazysetup}
\end{figure}

We test various values for the setup interval parameter. As shown in Fig.~\ref{lazysetup}, the prolongators $\mathbf{P}$ derived from the system matrices $\mathbf{A}$ in the first frame remain effective for subsequent frames. Even under aggressive settings where the interval is set up to $100$. In practice, we choose a relatively conservative interval $20$ in all our other experiments. Compared with the exact evaluation of $\mathbf{P}$ every frame which can take almost two-thirds of the total time, our setting reduces the setup time to only 2\% without any noticeable degradation.

\subsection{Near-kernel/near-nullspace Components}

A key input for the setup phase of our AMG is the near-kernel component. This notion relates to algebraic smoothness, which can be explained by the eigen decomposition \cite{Falgout2006-IntroAMG}. For residual equations $\mathbf{A}\mathbf{e}=\mathbf{r}$, where $\mathbf{A}$ is invertible,  $\mathbf{e}$ can be eigen-decomposed. The eigenvector corresponding to the smallest eigenvalue is an algebraically smooth error, named near-kernel/near-nullspace component $\mathbf{B}$. The near-kernel component, as its name implies, serves as an approximate solution to the homogeneous linear system $\mathbf{A}\mathbf{x} = \mathbf{0}$.  In aggregation-based AMG, the near-kernel component is critical for the effectiveness of the prolongator. Existing methods utilize all-ones vectors, rigid-body modes, or Adaptive Smoothed Aggregation (adaptive-SA) to design the near kernel. However, the all-ones vector does not perform well for non-Poisson equations, and identifying rigid-body modes within the dual system presents challenges. Adaptive-SA generates the near kernel using another preliminary AMG, which is widely applicable but incurs high costs. 

\begin{figure}[th]
    \centering
    \includegraphics[width=\linewidth]{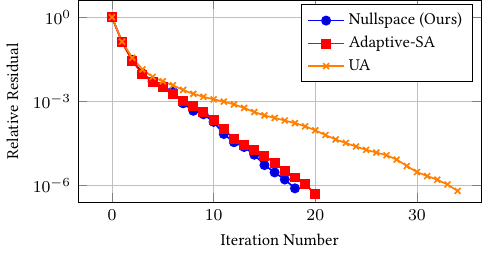}
\caption{Comparison between our improved near nullspace method and the adaptive SA in achieving $10^{-6}$ residual in the linear system. Our method only requires half of the iterations of UA and is similar to the Adaptive-SA.  The data are collected from the 270K-resolution bunny squash case. }
\label{fig:nullspace}
\Description{A line plot depicting the convergence curves of the improved near nullspace method and the adaptive SA.}
\end{figure}

Inspired by adaptive-SA, which uses AMG to bootstrap AMG, we propose a simplified approach that executes a few sweeps of iterative methods on the homogeneous equation to design the near-kernel. This simple method exhibits convergence similar to that of adaptive-SA (Fig.~\ref{fig:nullspace}). Notably, compared with UA, it only needs half the number of iterations of UA to reach the same convergence level. We find that 20 sweeps of GS on $\mathbf{A}\mathbf{x}=\mathbf{0}$ with $\mathbf{x}_0$ randomly sampled in \((0, max(|A_{ij}|))\) is good enough for most  cases. We repeat this six times to generate six distinct $\mathbf{B}$ for bootstrapping our AMG, because the solid motion consists of six rigid body modes.

\begin{figure}[tp]
    \centering
    \subfigure[]{\includegraphics[width=0.45\linewidth]{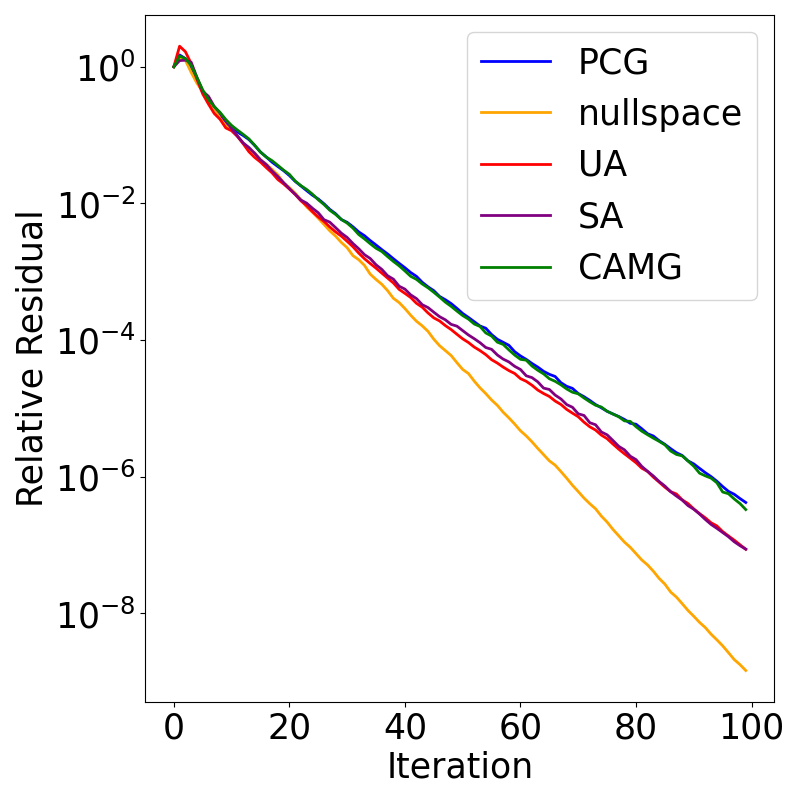}\label{fig:compare-setup-phases}}
    \subfigure[]{\includegraphics[width=0.45\linewidth]{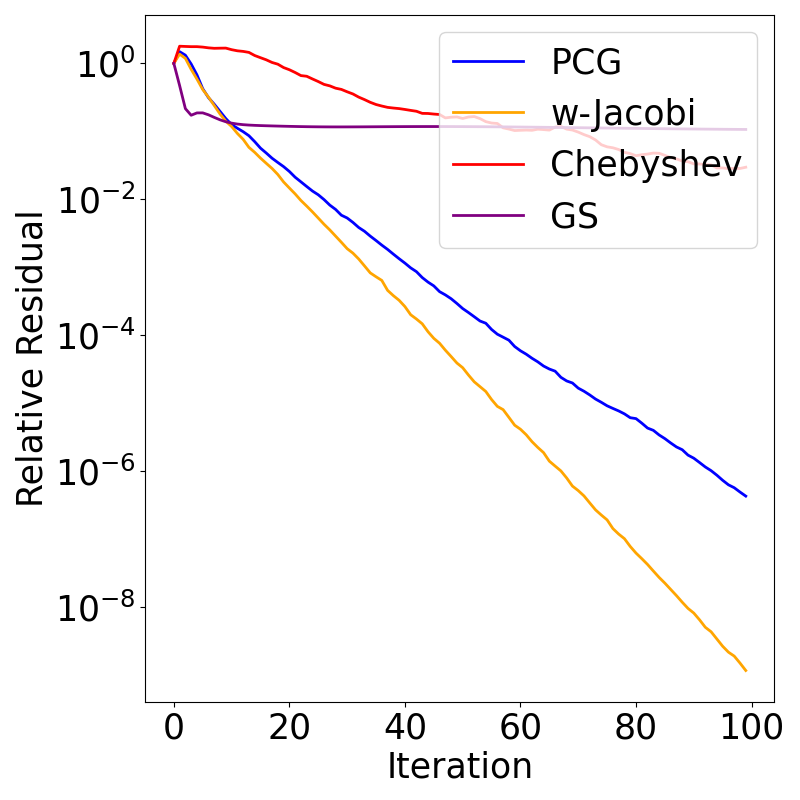} \label{fig:compare-smoothers}
    }
    \caption{\textbf{Comparison of linear solvers} for different AMG methods and different smoothers in convergence rate. a) Comparison of Different setup phases, the smoother uses weighted-Jacobi method.  It shows that the nullspace method (ours) is superior. b) Different smoothers, the setup phase uses the nullspace method. It shows that the weighted-Jacobi method is the best smoother for softbody cases. The data is collected from the bunny squashing case with 12K tetrahedrons (Fig.\ref{fig:bunny}). }
    
    \Description{Two line plots to compare the convergence curve of a) different setup method b) different smoothers. }
    \label{fig:compare-linearSolver}
\end{figure}

\subsection{Solving Phase}
With prolongators generated during the setup phase, we can proceed to the solving phase. In this phase, we first construct the coarse matrices for each level using the Galerkin principle:
\begin{equation}
   \mathbf{A}_{l+1} = \mathbf{P}_l^{\top} \mathbf{A}_{l} \mathbf{P}_l, \label{eq:Galerkin}
\end{equation} 
where $l$ denotes the level index.

Next, we utilize an MGPCG loop to solve the linear system. For each PCG iteration, one step of the V-cycle is performed to serve as a preconditioner. According to \citet{Stuben2001-ReviewAMG}, MGPCG is more robust and efficient than a standalone AMG solver.

\subsection{Smoother}\label{sub:smoother}
We implement three GPU-accelerated smoothers, including weighted Jacobi ($\omega$-Jacobi), parallel GS, and Chebyshev. To maintain symmetry in the multigrid V-cycle, identical pre- and postsmoothers are employed, each executing two sweeps of the iterative solver.  We found that the $\omega$-Jacobi has the best performance for softbody, and Chebyshev has the best performance for cloth (Fig.~\ref{fig:compare-linearSolver}).

The weighted Jacobi has an optimal $\omega_{\text{opt}}$ =  \( \frac{2}{\lambda_{\text{max}}(\mathbf{D}^{-1}\mathbf{A}) + \lambda_{\text{min}}(\mathbf{D}^{-1}\mathbf{A})}\). $\lambda_{\text{max}}$ is computed using the power method. Computing \(\lambda_{\text{min}}\) is much harder than \(\lambda_{\text{max}}\), so we approximate it with a user-defined estimate (e.g., \(0.1\)), sacrificing accuracy for practical efficiency. This approximation boosts performance by \(~24\%\) in the 850K-tetrahedron bunny case (Fig. \ref{fig:bunny}).

In particular, we find that both the weight of $\omega$-Jacobi and the Chebyshev polynomial coefficients can also be updated lazily during the AMG setup phase, and in this way we could reduce computational overhead without compromising convergence.

\section{Results}

\subsection{Comparision with Primal Space Methods}

\begin{figure}[t]
    \centering
    \includegraphics[]{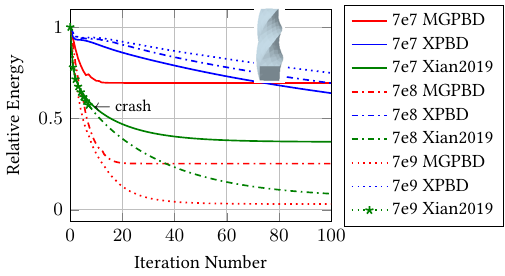}
    
    \caption{Relative energy of \textbf{Bar Twist} under various stiffness comapred with primal space method. Relative energy is normalized by dividing it by the energy before the loop.}
    \label{TwistBar}
    \Description{Line plot showing energy decrease over iterations for different stiffness levels.}
\end{figure}

As noted in \citet{Macklin2020-primal-dual}, the system matrix of the primal space method is 
\begin{equation}
\mathbf{A}_{primal} = \nabla \mathbf{C}^T \boldsymbol{\alpha}^{-1} \nabla \mathbf{C} + \mathbf{M} / \Delta t^2.
\end{equation} 
The high stiffness-to-weight ratio causes ill-conditioning in the primal space, leading to significant errors such as excessive stretching. Primal-space methods, e.g.~\cite{kimdeformables, Xian2019-GalerkinMG, Chen2024-VBD, Ruan2024MiNNIE}, struggle in high-stiffness ($10^9$ order) and high-resolution scenarios. Although dual formulations are generally insensitive to stiffness-to-weight ratios, existing tools, such as Houdini's XPBD \cite{Houdini}, introduce physically incorrect softening artifacts under these conditions. As illustrated in Fig. \ref{TwistBar}, a $5\times1\times1$ bar is twisted by $180^\circ$ and released. Our method outperforms both \citet{Xian2019-GalerkinMG}'s method (a state-of-the-art primal-space method utilizing multigrid techniques) and XPBD in high-stiffness scenarios. The parameters used are $E = 7\times10^7/10^8/10^9$, $\nu = 0.499$. Our method outperforms both the primal-space multigrid method and XPBD as the stiffness increases. At the highest stiffness levels, \citeauthor{Xian2019-GalerkinMG}’s method fails within the first $7$ iterations, XPBD stalls, while our method continues to exhibit a steady decrease in energy.

\begin{table*}[]
\caption{\textbf{Experiments setup and performance}. Ball cases (Fig.~\ref{fig:ball}) are to test whether the solver shows stiff results under different resolutions. Bunny squash are set to compare the performance. Cloth (Fig.~\ref{fig:cloth}) cases use unlimited iterations until they converge to 1e-4 to test the inextensibility of high-resolution meshes. Muscle-human (Fig.~\ref{fig:teaser}) and muscle-Monster (Fig.~\ref{fig:monster}) cases set maxiter=$20$ instead of using a time budget, so "timebudget" means time here. Beam cases are designed to test the stability under large $\Delta t$. Beam cases (Fig.~\ref{fig:beam}) are testing the different $\Delta t$. $C$ and $nl$ are operator complexity and number of layers of AMG of highest resolution cases, they are used to evaluate the complexity of AMG. }\label{tab:all-perf}
\begin{tabular}{lllllllllll}
\toprule
case name       & \#tets(\#edges)                     & \#verts          & time budget(s/frame) & $\Delta t$             & stiffness & C     & nl \\
\midrule
ball (Fig.~\ref{fig:ball})            & 1K/22K/99K                 & 0.4K/5.7K/24K    & 1s                  & 10ms           & 1e9   & 1.000 & 3  \\
bunny-squash (Fig.~\ref{fig:bunny})    & 5K/12K/270K/850K           & 1.3K/3K/60K/185K & 0.5/1/5/10s         & 10ms           & 1e9   & 1.002 & 3  \\
cloth   (Fig.~\ref{fig:cloth})          & 8K/32K/131K/524K  & 4K/16K/66K/263K  & maxiter=1e5         & 3ms            & 1e9   & 1.046 & 5  \\
beam   (Fig.~\ref{fig:beam})           & 2.9K                       & 0.8K             & 0.5s                & 10/20/30ms & 1e12   & 1.000 & 2  \\
muscle-human (Fig.~\ref{fig:teaser})    & 1667K                      & 526K             & 40.6s (maxiter=20)  & 3ms    & 1e9   & 1.001 & 4  \\
muscle-monster  (Fig.~\ref{fig:monster})  & 72K                        & 26K              & 2.3s (maxiter=20)   & 3ms    & 1e9   & 1.003 & 3   \\
bar-twist  (Fig.~\ref{TwistBar})  & 32K                        & 5K              & N/A   & 10ms    & 2.33e7/e8/e9   & 1.003 & 3   \\
collision (Fig.~\ref{fig:collision})  & 34K & 8K & 8s & 3ms & 1e9 &1.002 & 2\\
\bottomrule

\end{tabular}
\end{table*}

\subsection{Performance}

To evaluate our methods, we implement our MGPBD using CUDA and Taichi Language \cite{taichi-lang-course-2020}. All experiments are tested on a PC with an AMD 7950x CPU, NVIDIA GeForce RTX 4090 GPU, and 64GB host memory. We present various examples to demonstrate the efficacy of MGPBD. Timing, statistics, and material parameters are given in Tab. \ref{tab:all-perf}. To ensure a fair comparison, we allocate the same time budget for both MGPBD and XPBD in most cases. All XPBD cases utilize a Jacobi solver running on a GPU implemented in Taichi. 

\begin{figure}[t]
\centering
\includegraphics[]{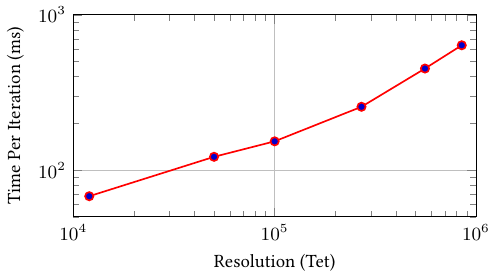}
\caption{Time per iteration to achieve $10^{-2}$ relative dual residual with respect to the resolution. It shows that the time per iteration is linearly scaled up with resolution, where $R^2=0.9978$ for linear regression. Data are from bunny-squash cases.}
\label{fig:TimeVsResolution}
\Description{Two line plots.}
\end{figure}

\begin{figure}[t] \centering    
\subfigure[] {
 \label{fig:time-dist}     
\includegraphics[width=0.5\linewidth]{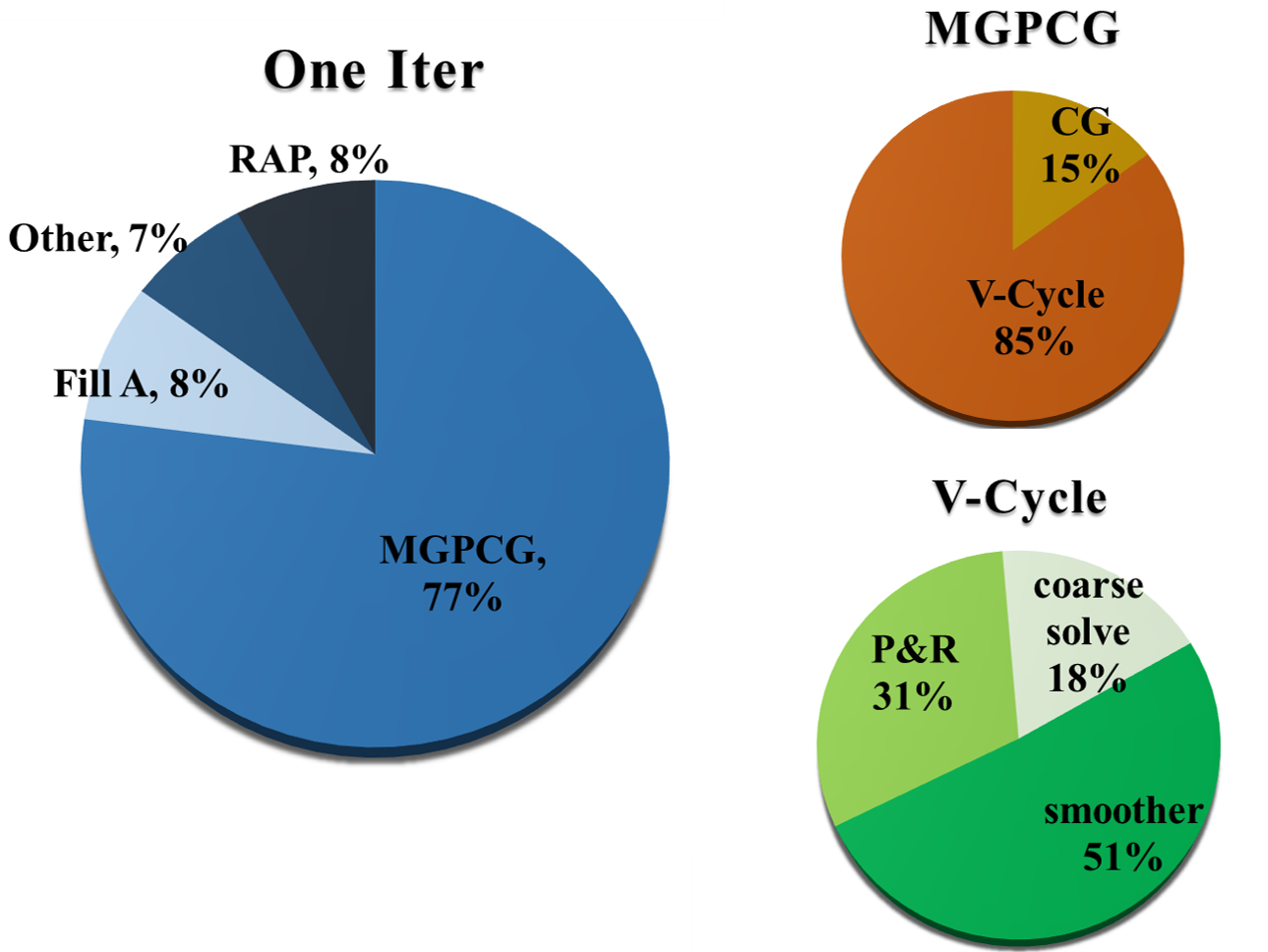}
}     
\subfigure[] { 
\centering
\label{fig:time-impl-compar}     
\includegraphics[width=0.45\linewidth]{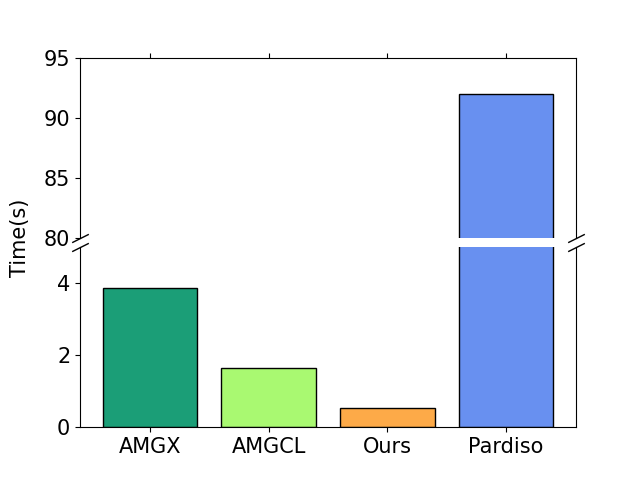}  
}    
\caption{Performance of AMG Implementation. a) Time distribution of each iteration.  b) Comparison with third libraries (One Iteration). Ours: $0.54$s; AMGX: $3.85$s; AMGCL: $1.63$s; Direct Solver (PARDISO): $91.98$s}    
\Description{a) A pie graph showing the distribution of time. b) A bar plot comparing the different methods.}
\label{fig:perf-impl}     
\end{figure}

There are already several highly optimized open-source AMG libraries available, such as NVIDIA AMGX on GPU~\cite{AMGX} and AMGCL on CPU backend~\cite{AMGCL-Demidov2020}. To ensure fair comparisons, we configured them with aggregation-type AMG and used the results of best performance. We also compare our results with Intel MKL's highly optimized PARDISO solver~\cite{PARDISO-SCHENK200169}, which employs a direct solver (Fig.~\ref{fig:perf-impl}). The results demonstrate the superiority of our MGPBD solver, achieving two or three orders of magnitude improvements.
Fig.~\ref{fig:time-dist} shows the time distribution of one iteration of MGPBD, where the MGPCG solve phase takes the largest portion. The major part of MGPCG is the V-cycle, which takes $85\%$ of an iteration. In the V-cycle, the most time-consuming part is the smoother. Fig.~\ref{fig:TimeVsResolution} shows the time taken to achieve a $10^{-2}$ relative dual residual. Our method shows a roughly linear scale-up performance with respect to the resolution ($R^2=0.9978$), which is an advantage of the AMG.

\subsection{As-Regid-As-Possible Constraints}
The As-Rigid-As-Possible (ARAP) constitutive model is used for softbody simulation and muscle scenarios. The formulation of ARAP constraints  \cite{Kim2020-DeformableTut} is:
\begin{equation}
    \mathbf{C}(\mathbf{x}_0,\mathbf{x}_1,\mathbf{x}_2,\mathbf{x}_3)=||\mathbf{F} - \mathbf{R}||_F ^2,
\end{equation}
where $\mathbf{F}$ is the deformation gradient, $\mathbf{x}_i(i=0,1,2,3)$ are vertices of four vertices of a tetrahedron, and $\mathbf{R}$ is the rotation matrix decomposed from $\mathbf{F}$. The detailed formulation of the constraint gradient refers to \citet{Kim2020-DeformableTut}. 
Experiment configurations are listed in Tab. \ref{tab:all-perf}. The compliance $\alpha=1/(\mu V_{tet})$, where $\mu$ is the second Lame parameter and $V_{tet}$ is the rest volume of the tetrahedron. For convenience, we refer to $\mu$ as stiffness because $V_{tet}$ varies.

\subsection{Scenes}
\paragraph{Bunny Squash}
Fig.~\ref{fig:bunny} shows bunnies at different resolutions (5K, 12K, 270K, 850K) recovering from a squashed state.  
Fig.~\ref{fig:Bunny-Squash-Residual} plots the relative dual residual curves versus iteration number (left) and time (right) respectively. Here the residual is normalized by dividing it by
the residual at the start of the loop.  It shows that XPBD stalls, showing a flat curve. In contrast, our method maintains continuous improvement, with the residual steadily decreasing.  For the 850K-tetrahedron bunny, XPBD diverges, while our method continues to show stable convergence.

\paragraph{Beam} 
Although \citet{Macklin2016-XPBD} claimed that XPBD addressed the timestep size and iteration numbers depended on the material stiffness, our experiments in Fig.~\ref{fig:beam} show that it still suffers the problems under a limited time budget in high stiffness scenarios. In Fig.~\ref{fig:beam}, three highly stiff beams bend, simulated  with time step sizes: $10$ms, $20$ms and $30$ms. Our method exhibits consistent simulation results and keeps the object stiff while the XPBD method shows unphysical softness with a large time step size.

\paragraph{Ball}
 Fig.~\ref{fig:ball} shows balls with different resolutions (1K, 22K, and 99K tetrahedra). Even though the same stiffness is given, the XPBD shows a resolution-dependent softness, but our method is able to keep the ball stiff because of better convergence.

\paragraph{Muscle}
To showcase the versatility and flexibility of our solver in complex scenarios, we implemented a muscle solver. We added various muscle-to-bone and muscle-to-muscle constraints by applying external constraints. Specifically, a few extra XPBD sweeps were conducted before the elasticity loop.  Our solver uses the same input and output as XPBD and thus is easy to mix with XPBD. We can blend it with off-the-shelf XPBD implementations such as the Vellum solver in Houdini~\cite{Houdini}, serving as a complement to those hard situations (high resolution and high stiffness) that traditional XPBD finds troublesome.

\paragraph{Monster}
 In the experiment (Fig.~\ref{fig:monster}), we simulated a Monster walking with a 72K-element tetrahedral mesh with high stiffness. XPBD simulation quickly becomes unstable and crashes, while the MGPBD remains stable, with an average of $2.3$ s/frame.

\paragraph{Musculoskeletal human}
 Fig.~\ref{fig:teaser} shows human muscles with $1.667$ million tets and $140$K external constraints performing a shrugging motion.  The average performance is $40.6$s/frame. This experiment shows that our MGPBD solver can stably and robustly simulate high-resolution and high-stiffness muscles.

\paragraph{Cloth}
Fig. \ref{fig:cloth} shows the hanging tests for cloth at different resolutions with distance constraints. We tested MGPBD and XPBD with a very large maxiter (1e5), and expected them to achieve $10^{-4}$ accuracy ($||b||<10^{-4}$). MGPBD shows better inextensibility, while XPBD encounters the stalling issue and cannot achieve the accuracy at high resolution.

\paragraph{Collision}
  We implemented a position-based collision response with a static signed distance field (SDF), which is called after the elasticity loop as a post-processing step. The collision response includes two steps: first, we move the penetrated particles to the collider's surface; second, we reflect the normal velocity of the penetrated particles. The results (as in Fig.~\ref{fig:collision}) show that our MGPBD method can support collisions for high-resolution deformable object simulations.
\section{Conclusion And Limitations}
We present an MGPBD method, a multigrid-enhanced version of XPBD, targeting convergence issues in high-resolution and high-stiffness simulations where XPBD struggles. MGPBD integrates directly into existing PBD/XPBD pipelines and uses a global AMG solver to achieve faster convergence. MGPBD performs better for demanding scenarios such as extremely stiff materials or finely detailed models.

Although our AMG framework demonstrates superior performance on static-topology systems, it still requires the construction of a sparse linear system matrix. In future work, we plan to explore a matrix-free AMG solver to eliminate the need for explicit matrix storage, thereby reducing memory usage and enabling support for dynamic topology changes such as tearing and fracturing. Additionally, the global system introduces extra oscillation, which requires damping. We also aim to incorporate more complex collision handling into our MGPBD framework.

\begin{acks}
This work was supported in part by the National Key R\&D Program of China (No.2023YFC3604500), Beijing Natural Science Foundation under Grant (No. 4252018), National Natural Science Foundation of China (L2324214), Beijing Science and Technology Project (No. Z231100005923039),the Postdoctoral Fellowship Program of CPSF under grant number GZC-20233375, the Guangxi Science and Technology Major Program (No. GuiKeAA24206017).
\end{acks}

\bibliographystyle{ACM-Reference-Format}
\bibliography{refs.bib}
\clearpage
\appendix
\begin{figure*}
    \centering
     \onecolumn
     \includegraphics[width=1.0\linewidth]{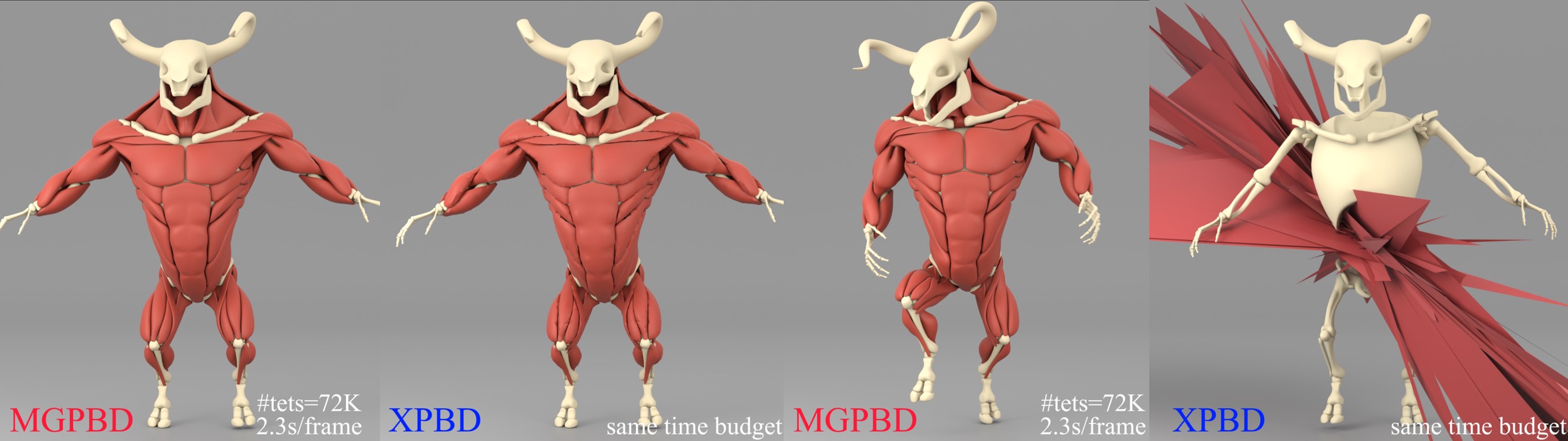}
    \caption{\textbf{Monster}. A monster animated with bones. The model has $72$K tets and $14$K external constraints connecting bones and muscles. The average speed of MGPBD is $2.3$ s/frame. The XPBD runs under time budget $2.3$ s/frame and failed. The stiffness is $1e9$.}
    \label{fig:monster}
    \Description{A monster is walking.}
\end{figure*}

\begin{figure*}
    \centering
    \includegraphics[width=1.0\linewidth]{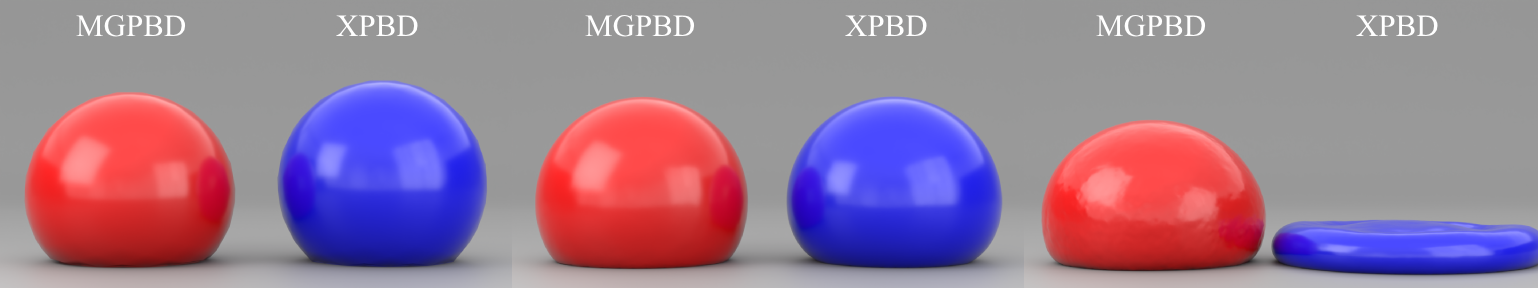}
    \caption{\textbf{Ball}. Different resolutions($1$K/$22$K/$99$K tetrahedrons from left to right) for a ball falling cases under the same time budget. MGPBD shows the stiff results in high-res cases.}
    \Description{A ball falls to the ground.}
    \label{fig:ball}
\end{figure*}

\begin{figure*}
    \centering
    \includegraphics[width=\the\linewidth]{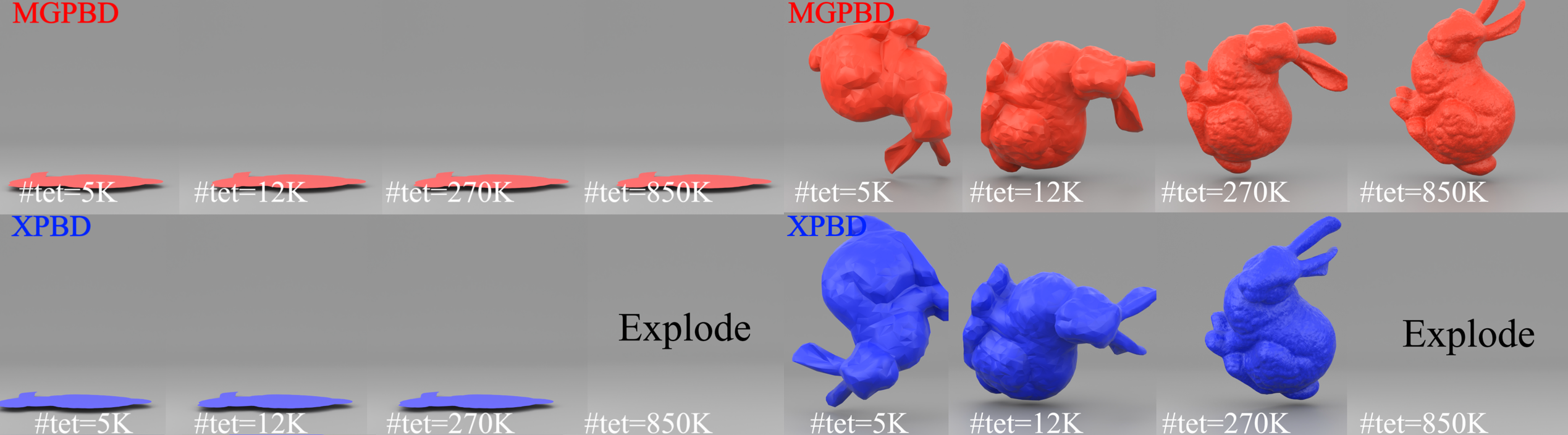}
    \caption{\textbf{Bunny Squash}. A bunny recovers from a squashed state under ARAP constraints. The MGPBD and XPBD are compared under the same time budget with different resolutions ($0.1$s for $5$K, $5$s for $270$K, and $10$s for $850$K). The $850$K XPBD crashes. $dt=10$ms, stiffness=$1e9$. XPBD runs in GPU.}
    \label{fig:bunny}
    \Description{A bunny recovers from the squashed flat state.}
\end{figure*}

\begin{figure*}
    \centering
    \includegraphics[width=\the\linewidth]{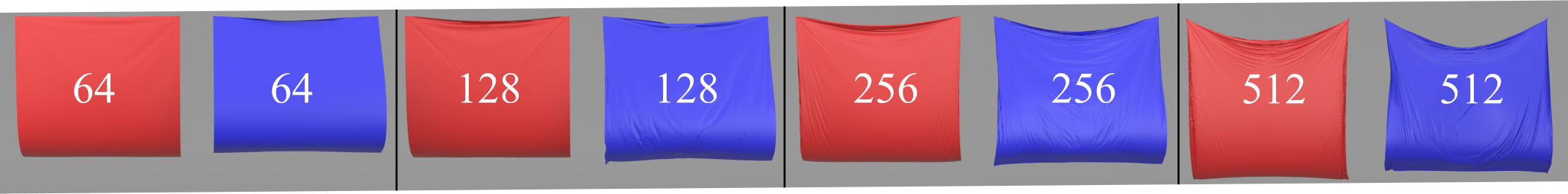}
    \caption{\textbf{Cloth}.  For mass-spring clothes with maxiter = 1e5, the simulation runs until the dual residual converges to $10^{-4}$ at various resolutions: \#vertices=$(N + 1)^2$ (N = $64/128/256/512$).  This case shows that MGPBD (red) has better inextensibility because it is free of stalling. XPBD (blue) encounters stalling issues at high resolutions even with such high iteration counts.}
    \label{fig:cloth}
    \Description{A cloth is hanging.}
\end{figure*}

\begin{figure*}
    \centering
    \begin{minipage}{0.48\textwidth}
        \centering
        \includegraphics[width=\linewidth]{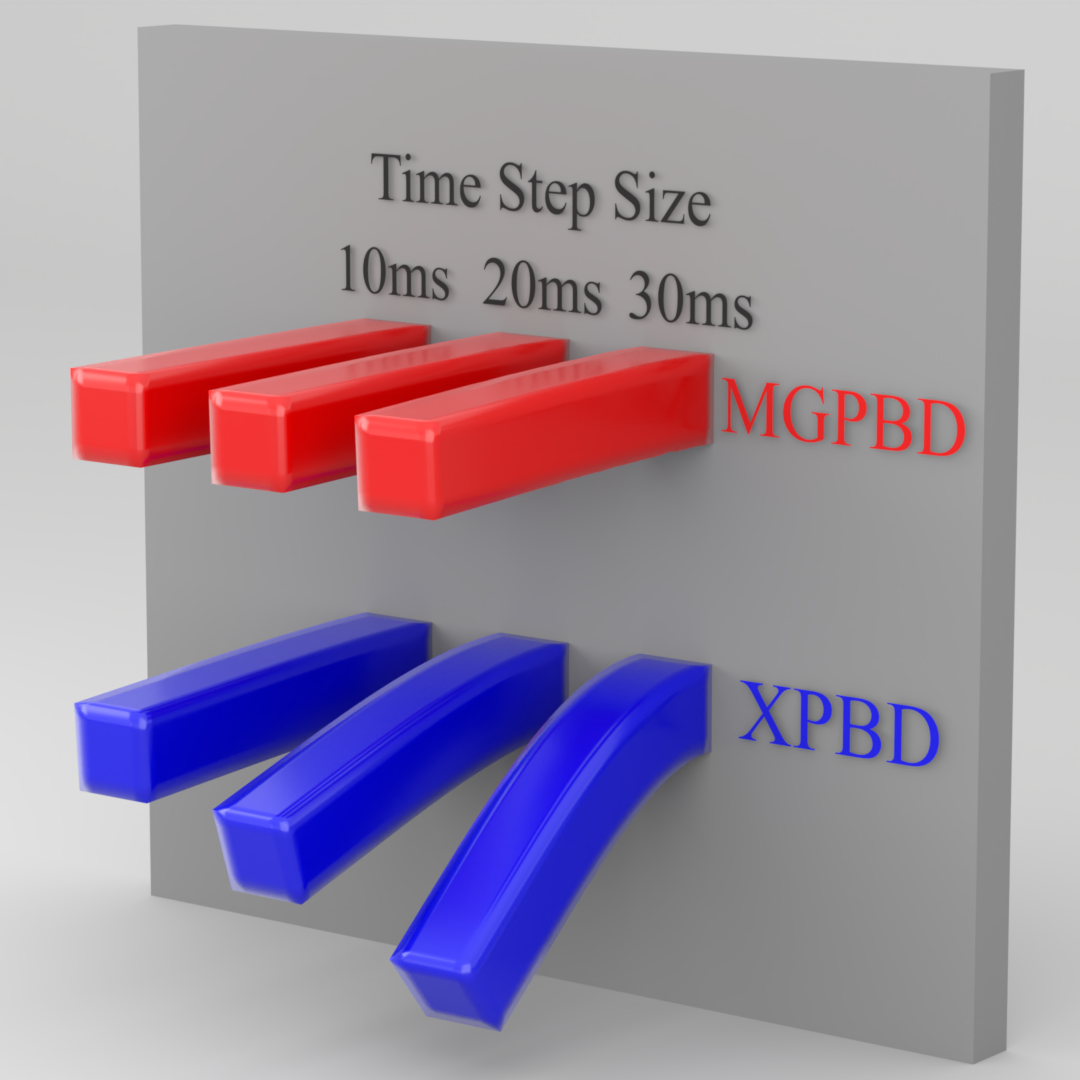}
        \caption{\textbf{Beam}. Different time step size $\Delta t$ ($10$ms/$20$ms/$30$ms) for a stiff(stiffness=$1e12$) beam under same time budget. MGPBD can withstand larger $\Delta t$ than XPBD.}
        \label{fig:beam}
    \end{minipage}
    \hfill
    \begin{minipage}{0.48\textwidth}
        \centering
        \includegraphics[width=\linewidth]{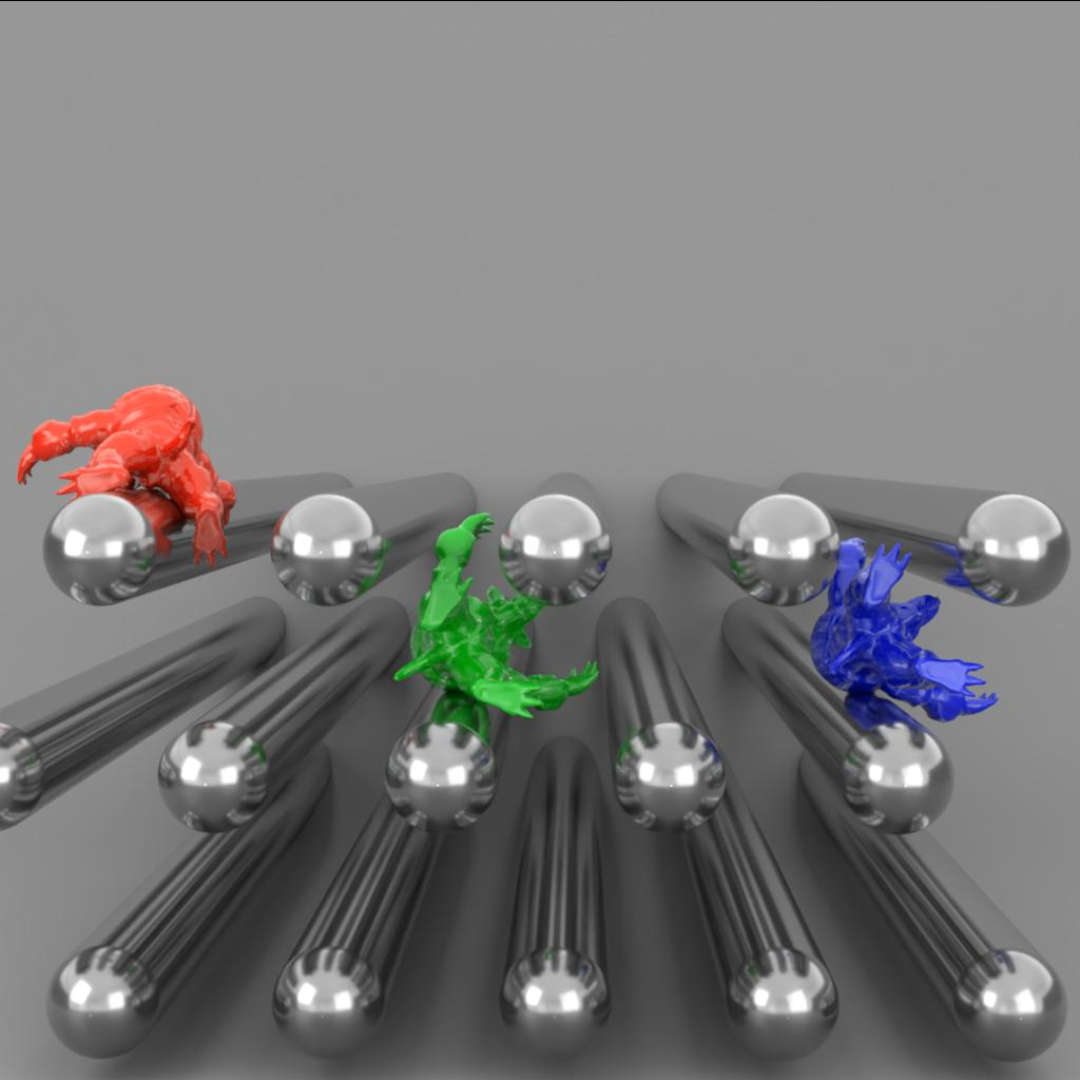}
        \caption{\textbf{Collision}. Three armadillos collided with $20$ cylinders defined in static SDF. A position based collision response is adopted after the MGPBD elasticity loop.}
        \label{fig:collision}
    \end{minipage}
    \Description{Three armadillos fall from sky and collide with many pillars.}
\end{figure*}

\begin{figure*}[thb]
    \centering
    \includegraphics[]{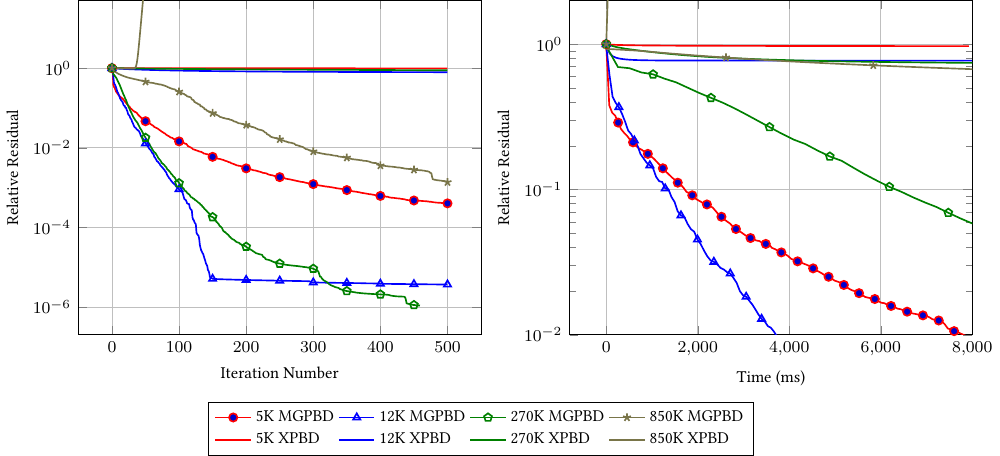}
    \caption{\textbf{Residual of Bunny Squash at various resolutions.} Relative dual residual versus iteration number (the left subfigure) and time (the right subfigure) in the bunny squash case. Our MGPBD solver avoids stalling, whereas XPBD stalls at all resolutions, and even crashes when the number of tetrahedrons reaches 850K. In the right subfigure, markers are spaced every $5$ iterations to compare per-iteration cost. The relative residual is normalized by dividing it by the residual at the start of the loop. }
    \label{fig:Bunny-Squash-Residual}
    \Description{Two line plots.}
\end{figure*}

\end{document}